\documentstyle[prd,aps,tighten,preprint]{revtex}
\begin{document}

\def\ii{\'{\i}}
\def\bi{\bigskip}
\def\be{\begin{equation}}
\def\en{\end{equation}}
\def\bq{\begin{eqnarray}}
\def\eq{\end{eqnarray}}
\def\noi{\noindent}
\title{\bf ON SCALE INVARIANCE AND ANOMALIES IN QUANTUM MECHANICS\footnote{Work supported by CONACyT under contract 39798-E}}
\author{A. Cabo\footnote{On leave of absence Instituto de Cibern\'etica, Matem\'aticas y F\ii sica, Calle E No. 309 Esq. 15 Vedado, La Habana, Cuba}, J.L. Lucio and H. Mercado}
\address{Instituto de F\ii sica, Universidad de Guanajuato \\
Apartado Postal E-143, Le\'on, Gto., M\'exico}
\maketitle
\noindent

\vspace{2.1cm}

\begin{abstract}
\setlength{\baselineskip}{.5cm}
We re-consider the quantum mechanics of scale invariant potentials in two 
dimensions. The breaking of scale invariance by quantum effects is analyzed by
the explicit evaluation of the phase shift and the self-adjoint extension 
method.  We argue that the breaking of scale invariance reported in the 
li\-te\-ra\-tu\-re for the $\delta$({\bf r}) potential, is an example of 
explicit and not an anomaly or quantum mechanical symmetry breaking.
\end{abstract}

\setlength{\baselineskip}{1\baselineskip}

\newpage

\noi {\Large\bf INTRODUCCION}

\bi
\bi

Symmetries play a central role in the description of physical systems. Well 
known exam\-ples [1] are the space-time symmetries (space and time 
homogeneity, 
{\it i.e.} invariance of the system under space and time translations) which 
are the basis for the energy and momentum conservation laws we learn to use in 
elementary classical  physics courses. The link between symmetry properties of
a system and conservation laws is provided by Noether's theorem [2] which 
asserts that associated to each transformation that leaves invariant the 
action there exist a conserved quantity.  The properties of the system can, in 
general, be obtained in terms of such  conserved quantities without completely
solving the equations of motion that describe the system. Besides the 
space-time symmetries there may exist ``internal" symmetries which are 
related to the conservation of quantities such as the electric charge. In 
fact, symmetry requirements are enough to fix the way electric charges 
interact, thus the question of what are the interactions that ocurr in nature
is traded by the more fundamental question: what are the symmetries of nature?

\bi

In the conventional approach, theories are formulated at the classical level 
and latter quantized according to a well established procedure [3]. Usually 
the symmetries survive the process of quantization ensuring thus the validity 
of conservation laws at the quantum level. Remember also that although the 
electromagnetic and the gravitational are long range interactions which may be
deal with at the classical level, the weak and strong are very short range 
interactions and inevitably require a quantum treatment. If we want to 
describe the interactions in terms of symmetries, we have to make sure that 
the symmetry is valid at the quantum level. {\bf Anomalies} occur when the 
symmetry is destroyed by quantization, a phenomenon originally identified in 
quantum field theory and recently analyzed in the context of quantum mechanics
[4]. In quantum field theory anomalies may have phenomenological consequences 
(when the anomaly is associated to a global symmetry) or render inconsistent 
the theory (if the anomaly is related to a local symmetry), this has motivated
the interest in anomaly free theories and the search for a deeper 
understanding of this symmetry breaking mechanism. Anomalies have not raised 
much interest in the framework of quantum mechanics, in fact only a few 
examples [5] have been analyzed in detail, although the possibility that they
are related to geometrical phases has been advanced in this context [6].

\bi

Systems invariant under {\bf dilation} of the space-time coordinates ($x \to 
\rho^{-1/2} ~x$ with $\rho$ an arbitrary dimensionles parameter) are refereed 
as scale invariant, a characteristic of systems not depending on dimensional 
parameters. In quantum field theory this symmetry is destroyed by quantum 
effects and a similar conclusion has been obtained in quantum mechanics [7]. 
In nonrelativistic physics, where the kinetic part of the Hamiltonian is 
proportional to {\bf p}$^2$, scale invariant systems are described by 
potentials 
such that $V(\rho^{-1/2} ${\bf r}) = $\rho V(r)$ (we say the potential and the 
Hamiltonian scale as $\rho$). The invariance of the system under this 
transformation follows from the fact that modifications of the Hamiltonian by 
an overall factor does not affect the equation of motion. Examples of 
mechanical scale invariant systems are the $1/r^2$ and, in two dimensions, the 
delta potential $\delta^2$ ({\bf r}). Scattering off the $1/r^2$ potential is 
exactly solvable, it yields an energy independent phase shift which has to be 
understood as a signal of scale invariance at the quantum level, {\it i.e.} 
for this system the symmetry survives quantization. This is a clean example 
where the methods to be used in more involved analysis can be tested. 

\bi

The interest in the delta potential arose from the study of the $\lambda\varphi^4$ theory, which is non trivial perturbatively, but suspected to be non 
interacting. Beg and Furlong [8] considered the non-relativistic limit of the
$\lambda \varphi^4$ ---which results in the quantum mechanics of the delta 
potential--- in order to get some insight into the behavior of the full 
theory. They concluded that for a finite, {\bf unrenormalized} coupling 
constant, a trivial S matrix is obtained. Notice that we advocated 
renormalization, a subtraction procedure required in quantum
field theory where singularities associated to short distances (ultraviolet 
divergences) have to be removed by renormalizing the Lagrangian, which amounts
to redefine the fields, the coupling and the mass. The strong singularity of 
the delta potential, and its ``contact" nature, suggest that regularization is
required in order to properly define the quantum mechanics of that system. In 
three dimensions, and also in two [9], it is possible to get non-trivial 
dynamics at the price of renormalizing the interaction. The two dimensional 
case has the further interest of scale invariance at the classical level and 
the possibility it  offers to study the survival of this symmetry to the 
quantization process. For that reason the two dimensional delta potential has 
been considered a pedagogical laboratory where field theoretical concepts such
as renormalization [10], renormalization group equation [11], anomalies [4,7] 
and dimensional transmutation [12] can be studied. 

\bi

The regularization procedure used to treat singular potential in quantum 
mechanics must posses several features; in particular it should preserve the 
symmetries, otherwise the results obtained for physical observable are 
meaningless as they do not reflect the properties of the original system. It 
is possible to argue that the regularization and renormalization are part of 
the quantization procedure, however the breaking of scale invariance {\bf is 
not} an intrinsic characteristic of the regularization of the delta potential;
it is indeed possible to envisage a symmetry preserving regularization 
procedure. This point has been overlooked in previous analysis, which define 
the potential in terms of a distribution sharing some properties of Dirac's 
delta, but not scale invariance [10]. In this paper we introduce a 
distribution with the adequate scaling  properties and explicitly work out the
scattering problem  to evaluate the phase shift (using Green functions and 
solving exactly the corresponding Schr\"odinguer Equation). The distinctive 
feature of this approach is that it requires regularization but not 
renormalization leading thus, in agreement with Beg and Furlong [8] and 
Jackiw [9], to a trivial S matrix. Thus, we conclude that the breaking of 
scale invariance reported in the literature is  not an anomaly but an 
explicit breaking.  

\bi

In classical mechanics the Poisson bracket of  Noether's charge with a 
dynamical variable yields the variation of such a variable under the symmetry 
transformation, for that reason the charge is called the symmetry generator. 
In quantum mechanics the Poisson bracket is replaced by the commutator, with 
the further constraint that the generator must be Hermitean (which may require
solving ordering ambiguities to avoid inconsistencies). This quantum 
mechanical charge is then used to build the unitary operator which carries the
corresponding transformation in Hilbert space. Cocycles are phases which 
appear as a necessary generalization of the group representation theory in 
quantum mechanics when the action, but not the lagrangian, is invariant under 
a symmetry transformation [13]. Usually one is not faced with cocycles because
they are trivial, {\it i.e.} it is possible to redefine the wave function 
and the operators to avoid them. If the cocyle appears and it is non trivial, 
this signals a symmetry breaking by quantum effects. Originally our interest 
in the delta potential was to show that cocycles is an alternative framework 
to analyze the breaking of scale invariance by quantum effects. We have been 
unable to reach our goal due to complications arising from the non-point 
nature of the scale transformations.

\bi

An alternative way to test for a symmetry after quantization is the 
self-adjoint extension. In the particular case of scale invariance and the 
delta potential, the method  relies on the behavior around the origin of the 
wave function and the fact that this must be invariant under a symmetry 
transformation, {\it i.e.} the relation between $\psi(0)$ and $\psi'(0)$ can 
not change by the action of the symmetry generator. This behavior at the 
origin is the bridge between the self-adjoint property of the Hamiltonian and 
the symmetry properties of the system. The reason is that the Hamiltonian for 
a free particle in two dimensions is not self-adjoint and that means that only 
certain class of functions -those having at the origin derivative proportional
to the function it self- are acceptable as solutions for this problem [16]. 
Different self-adjoint extensions correspond to different relations $\psi(0)=
\lambda\psi'(0) (\lambda$ is called the self-adjoint extension parameter) and 
$\lambda$ is related to the phase shift, therefore different $\lambda's$ 
correspond to different potentials. In the main text we discuss in detail the 
self-adjoint extension for the $\delta$({\bf r}) potential. We show that 
the result of this analysis is completely consistent with that obtained by the 
explicit evaluation of the phase shift. We remark that this approach provides 
information about the symmetries of the system at the quantum level, but not 
on the nature of the symmetry breaking mechanism, in particular this is not a 
criteria for the existence of anomalies. 

\bi

Summarizing, at the classical level the two dimensional delta potential 
defines a system for which the action (but not the Lagrangian, which implies 
the appearance of cocycles) is invariant under scale transformations. For the 
same system in quantum mechanics we face the following alternative: {\it a)} 
we deal with a finite, unrenormalized coupling constant, a trivial S matrix 
and clearly scale invariance {\it b)} after renormalization of the coupling 
constant, the delta potential leads to a non trivial S matrix and an energy 
dependent phase shift indicating the breaking of scale invariance. ~~Besides 
~~the ~explicit ~calculation of the energy dependent phase shift, Jackiw 
confirmed [9] these results in terms of the self adjoint 
extension. Thus the breaking of scale invariance is out of the question, 
however and that is the point we address in this paper, the nature of the 
breaking is not evident. In fact, and contrary to some claims in the 
literature, we argue that this is an example of explicit and not an anomaly or
quantum mechanical symmetry breaking. 

\bi
\bi

\noi {\Large\bf SCALE INVARIANCE.}

\bi

In the following we restrict our attention to two dimensions. The vector 
components are labeled by Latin indices {\it i} = 1,2 .  The finite scale 
transformation are defined by:

\bq
t{\buildrel T \over \longrightarrow} t^\prime =\rho t & ~~x_i (t) {\buildrel 
T\over\longrightarrow} ~~x^\prime_i (t^\prime)=\rho^{-1/2}x_i (\rho t), \nonumber \\
& p_i (t) {\buildrel T\over\longrightarrow} ~~p^\prime_i (t^\prime)=\rho^{1/2} p_i (\rho t) .
\eq

We consider the $1/r^2$ and the $\delta^2${\bf (r)} potential which have the 
same properties under scale transformations $(U(r){\buildrel T\over\longrightarrow} U(\rho^{-1/2} r) = \rho U(r)):$

$$
\frac{1}{r^2}{\buildrel T\over\longrightarrow} \frac{1}{r^2}=\frac{\rho}{r^2},
$$ 
$$
\delta^2 ({\bf r}){\buildrel T\over\longrightarrow} \delta^2 ({\bf r}\,^\prime)
=\rho \delta^2 ({\bf r}).$$

A conventional approach to regularize the $\delta ({\bf r})$ potential amounts
to  the replacement [10]:

\be
\delta^2 (\bf r) \longrightarrow \left\{ \begin{array}{ll} \frac{v (a)}{\pi a^2} & r \leq a, \\
\mbox{}\\
0 & r > a .
\end{array}
\right. 
\en

\noi In the following we consider two possibilities for $v(a)$:

$$v (a) = \frac{2\pi}{\ln (a/a_0) + \gamma}, \eqno (3a)$$
$$v(a) = \upsilon = {\rm constant}. \eqno (3b)$$ 

\noi The first (3a), has been used in the literature [10] in order to obtain a 
non trivial S matrix. Below we argue that (3b) leads to a more appropriated
regularization of the $\delta (\bf r)$ potential. 

\bi

\noi Under scale transformation the ``regularized potential" transforms 
according  to:

\[ U (r^\prime) = \left\{ \begin{array}{lll} \frac{v (a)}{\pi a^2} & r^\prime 
=\frac{r}{\sqrt{\rho}}\leq a\to r\leq\sqrt{\rho} ~a \equiv a^\prime, \\
0 & r^\prime =\frac{r}{\sqrt{\rho}} >a \to r > \sqrt{\rho} ~a\equiv a^\prime , 
\end{array}
\right. \]

\noi which can be rewritten as:

\[ U (r^\prime) = \left\{ \begin{array}{lll} \frac{\rho v(a^\prime/\sqrt{\rho})}{\pi a^{\prime 2}} & r \leq a^\prime , \\
0 & r > a^\prime .
\end{array}
\right. \]

\bi

\noi In the  $a \to 0$ limit (2,3) describes a ``contact" or ``zero range"
potential, however {\bf it does not} share with the $\delta$({\bf r}) the 
properties under scale transformations, unless $v (a)$ is $a$ independent. 
Therefore, by using the distribution (2, 3a), and before doing any quantum 
mechanics, one introduces an {\bf explicit} breaking of scale invariance. On 
the other hand (3b) defines a family of probe functions appropriated for a 
mathematical definition of the $\delta^2 ({\bf r})$, and also ensures the 
adequate scale transformation properties of the regularized $\delta^2$
({\bf r}) potential. 

\bi

Notice that for infinitesimal transformations $(\rho \equiv 1+\epsilon + O 
(\epsilon^2))$, scaling (Eq.(1)) involve the velocities, {\it i.e.} these are 
non point transformations which may render difficult its implementation in 
quantum mechanics. Both for the $1/r^2$ and the $\delta^2$({\bf r}) 
potentials, the variation of the Lagrangian under the infinitesimal scale 
transformation is the time derivative of the Lagrangian $\delta {\cal L} = 
\frac{d{\cal L}}{dt}$. This non vanishing variation ensures the appearance of 
cocycles once the transformation is implemented at the quantum level [13]. The
charge associated to this symmetry is obtained through Noether's theorem [2]

$${\bf D} = \sum_i \frac{\partial{\cal L}}{\partial\dot x_i} \delta x_i -{\cal L} ={\cal H} t-\frac{1}{2} {\bf p} \cdot {\bf r} .$$

It is straightforward to check that both at the classical and quantum level 
{\bf D} generates the infinitesimal scale transformations:

\bq
\{ x_i, {\bf D}\} = \dot x_i t- \frac{1}{2} x_i & \{ \dot x_i, {\bf D}\} = \ddot x_i t + \frac{1}{2} \dot x_i , \nonumber \\
-\left[ x_i, {\bf D} \right] = \dot x_i t-\frac{1}{2} x_i & \left[ \dot x_i, {\bf D} \right] = \ddot x_i t + \frac{1}{2} \dot x_i . \nonumber
\eq

\bi
\bi

\noi {\Large\bf EXACT SOLUTIONS.}

\bi
\bi

Let us consider the scattering, in two dimensions, of a particle of mass $m$ by
a central potential $U(r)$. The hamiltonian of the system is 

$${\cal H} = \frac{{\bf p^2}}{2in}+U(r). $$

\noi For central potentials and in two dimensions, the angular momentum 
eigenfunctions $e^{i\ell\theta}$ are used to reduce the Schr\"odinguer 
equation 

$$\psi ({\bf r}) = \varphi (r) e^{i\ell\theta} ,$$

\noi the radial wave function $\varphi (r)$  is  obtained as a solution to

\setcounter{equation}{3}
\be
\bigg(\frac{d^2}{dr^2} + \,\frac{1}{r} \,\frac{d}{dr} -\,\frac{\nu^2}{r^2} + \kappa^2\bigg) \varphi (r) = 0 ,
\en

\noi where 

\[ \nu^2 (\ell) = \ell^2 + 2m \lambda , \quad\quad \kappa^2= k^2=2mE , \quad\quad {\rm if} \,\, U(r)=\lambda/r^2 \]
\begin{equation}
\nu^2 (\ell)  = \ell^2 , \quad \kappa^2_{\delta} =
 \left\{
  \begin{array}{lll}
  k^2 = 2mE , & r > a  &{\it if ~U(r) ~is} \\
  \kappa^2 =k^2-2m \frac{v(a)}{\pi a^2}, & r\leq a,
  & {\it given ~by} ~(2,3). 
 \end{array}
 \right.
\end{equation}

\bi

For fractional $\nu$, two independent solutions [14] are the first $J_\nu 
(\kappa r)$ and second class $Y_\nu (\kappa r)$ Bessel functions. For $\nu$ 
integer and $r < a$ two independent solutions are the modified Bessel 
functions $I_\nu (\kappa r)$ and $K_\nu (\kappa r)$. 

\bi
\bi

\noi {\Large\bf $1/r^2$ potential.}

\bi 

The $Y_\nu (\kappa r)$ function is discarded due to its singular behavior at 
the origin, therefore the physically acceptable radial wave function is given 
by:

\be
\psi_\ell (r,\theta) = N J_{\nu(\ell)} (k r) e^{i\ell\theta}, \qquad\qquad 
\ell = -\infty , \ldots \infty .
\en

\noi On the other hand, the solution corresponding to an outgoing free wave 
has the asymptotic behavior:

\be
\psi^A_\ell (r,\theta) \to \frac{e^{i\ell\theta}}{\sqrt{r}} \cos (kr-\frac{\ell \pi}{2} - \frac{\pi}{4} + \delta_\ell) .
\en

\noi The scattering phase shift is obtained by comparing the asymptotic 
expansions of (6,7) (the coefficient N is chosen so that, for $U(r) = 0$ the 
solutions coincide everywhere).

\be
\delta_\ell (k) =\frac{\pi}{2} (\ell - \sqrt{\ell^2 + 2m\lambda}).
\en

\noi This expression shows that phase shift is energy $(k)$ independent, which
is the signature for  scale invariance.

\bi

One may question this derivation due to the long range behavior of the $1/r^2$
potential. Further details about this problem are presented in the following 
section, where the phase shift is calculated using the scattering wave 
function.

\bi
\bi

\noi {\Large\bf $\delta^2 ({\bf r})$  potential}

\bi    

Since this potential is not  a function but  a distribution, we consider  
the set of spherically symmetric potentials vanishing outside a circle of 
radius {\it a} defined in (3a,b). The first alternative (3a) describes, in the
$a \to 0$ limit, the contact potential leading to non-trivial scattering of 
the so called renormalized $\delta ({\bf r})$ potential. This is the case 
discussed in the literature [10] that results in an energy dependent phase 
shift indicating the breaking of scale invariance, which has been identified 
by some authors as an anomaly in quantum mechanics [4]. On the other hand, 
when $v (a)=$ constant, the distribution defined by (2,3b) has, in the $a\to 0$
limit, the same scaling behavior than the $\delta^2({\bf r})$.

\bi

Outside the potential well, {\it i.e.}  for $r > a$, we write the solution as:

\be
\psi^e_\ell (r,\theta)=(b_\ell J_\ell (kr)+c_\ell Y_\ell (kr)) e^{i\ell\theta}.
\en

\bi

\noi Notice, for future reference, that comparing (7,9) in the asymptotic 
region allow us to conclude $b_\ell =\cos \delta_\ell, c_\ell = - sin \delta_\ell .$
 
\bi

On the other hand, in the internal region the solution is a linear combination
of the modified $I$ and $K$ Bessel functions. Again the $K_\nu (\kappa r)$ 
function is discarded due to its singular behavior at the origin. Thus, for 
$r < a$ the solution is given by:

\be
\psi^i_\ell (r,\theta)= d_\ell I_\ell (\kappa r) e^{i\ell\theta} ,
\en

\bi

\noi the coefficients $b_\ell, c_\ell$  entering in the external solution, for
each value of the angular momentum $l$, can be expressed in terms of the 
$d_\ell$ coefficients of the internal solution by matching the wave functions 
and their derivatives at $r = a$, thus

\bq
d_\ell &= \frac{1}{D} \bigg( \bigg(\frac{k}{\kappa} \bigg) J_\ell (ka) Y^\prime_\ell (ka) - J^\prime_\ell (ka) Y_\ell (ka)\bigg) b_\ell , \nonumber \\
c_\ell &= \frac{1}{D} (J_\ell (ka) I^\prime_\ell (\kappa a) - (\frac{k}{\kappa}) J^\prime_\ell (ka) I_\ell (\kappa a)) b_\ell , \nonumber
\eq

\noi where

$$D=\bigg(\frac{k}{\kappa}\bigg) Y^\prime_\ell (ka) I_\ell (\kappa a) - Y_\ell (k a) I^\prime_\ell (\kappa a) .$$

\bi

Let us first consider non-vanishing values of the angular momentum $l\not= 0$. 
Using the behavior of the Bessel functions for small arguments it is easily 
seen that both coefficients vanish in the $a \to 0$ limit. Thus, the phase 
shift and the scattering cross section vanish for non-zero values of the 
angular momentum, implying that the zero range potentials we are considering 
can only produce S wave scattering.

\bi

For zero angular momentum $l=0$, we consider two different cases.  First we 
take for $v(a)$ the expression (3a) used in the literature [10]. In this case 
$c_0, d_0$ take the following non-zero values in the limit of vanishing 
radius $a \to 0$: 

$$d_0=-\frac{\ln (\frac{a}{a_0})+\gamma}{\ln (\frac{k a_0}{2})}b_0,\eqno(10a)$$
$$c_0 =-\frac{\pi}{2} \frac{b_0}{\ln (\frac{ka_0}{2})}, \eqno (10b)$$

\noi from which we obtain the phase shift (see comment beneath eq. (9))

\be
tan ~\delta_0 (k) =-\frac{c_0}{b_0} =\frac{\pi}{2} \bigg(\ln \bigg( \frac{k a_0}{2} \bigg)\bigg)^{-1} .
\en

A second possibility we suggested in section two, is to consider $v(a)=v=$ 
constant. In this case the $c_0$ coefficient, and also the phase shift, 
vanish in the $a \to 0$ limit. The vanishing of the wave function at the 
origin in this case is also worth noticing.

\bi

>From this exercise we obtain the following conclusions: 

\begin{quote}
If the delta potential is treated in terms of the distribution (2,3a), which 
according to our reasoning breaks scale invariance at the classical level, 
then we get an energy dependent phase shift indicating the breaking of scale 
invariance in quantum mechanics.

\bi

If instead we use the distribution (2,3b), then we get a vanishing phase 
shift, and scale invariance is an exact symmetry both at the classical and 
quantum level.

\bi

The triviality of the $S$ matrix is not related to the strength of the 
potential. This is concluded by comparing the behaviour at the origin of 
(2,3a) and (2,3b).
\end{quote}

\bi
\bi

\noi {\Large\bf SCATTERING  WAVE  FUNCTIONS.}

\bi

An alternative derivation of the results of the last section can be obtained 
in terms of the scattering wave functions. We start with the Schr\"odinguer 
equation and its formal solution written in terms of the Green function:

\be
\psi^+ ({\bf r})= e^{i\vec k \cdot{\bf r}}-\int G({\bf r}-{\bf r}\,^\prime) u({\bf r}^\prime) \psi^+ ({\bf r}\,^\prime) d^2 {\bf r}^\prime ,
\en

\noi where the Green function is defined by the differential equation: 

$$(\nabla^2 + k^2)G({\bf r} - {\bf r}\,^\prime) =- \delta ({\bf r}-{\bf r}\,^\prime), $$

\noi plus the condition that the wave function $\psi^+({\bf r})$ describes a 
plane wave plus an outgoing ``spherical" wave. For ${\bf r}-{\bf r}\,^\prime 
\not=0$, the radial equation reduces to the Bessel equation (4), and the 
outgoing spherical condition selects ${\cal H}_0$ as the solution.

$$G({\bf r} -{\bf r}\,^\prime)=\frac{i}{4}{\cal H}_0 (k |{\bf r}-{\bf r}\,^\prime|) .$$

\noi The normalization $(c=-\frac{i}{4})$ is fixed by the strength of the 
source at the origin 

$$c \int\limits_{V\to 0} d^2 r^\prime (\nabla^2 + k^2) {\cal H}_0 (k r^\prime)
=-\int\limits_{V\to 0} d^2 r^\prime \delta (\vec r\,^\prime)=- 1$$

Returning to the scattering wave function, it is customary to show that 
asymptotically $\psi^+ ({\bf r})$ describes a plane wave plus an outgoing 
``spherical" wave, defining in this way the scattering amplitude $f(\theta)$:

\be
\psi^+(\vec r)=e^{i{\bf k\cdot r}}+f(\theta) \frac{e^{i(kr+\frac{\pi}{4})}}{\sqrt{r}}
\en

For short range potentials [15] ({\it i.e.} if the potential $U(r$) vanishes 
exactly for $r > a$ for some finite {\it a} it is enough to approximate the 
argument of the Green function $G({\bf r - r\,^\prime})$ by carrying aut an 
expansion in $(r^\prime/r)$ (remember that the vector ${\bf r}$ is understood 
to be directed towards the observation point at which the wave function is 
evaluated (${\bf r}\to \infty)$ whereas the region that give rise to a 
nonvanishing contribution $({\bf r\,^\prime})$ is limited in space for a 
finite range potential) and considering the asymptotic expansion $\lim\limits_{r\to\infty} {\cal H}_0 (kr -k \frac{{\bf r \cdot\vec r\,^\prime}}{r})$. The 
same approximation is justified at length in Quantum Mechanics textbooks for 
the one dimensional case (to which our problem is reduced by the 
transformation $\varphi (r)=\tilde \varphi (r)/\sqrt{r}$) and for potentials 
falling off faster than $1/r$. Thus, both for the $1/r^2$  and $\delta^2$({\bf r}) potentials, (12) is written as:

\be
\psi^+ ({\bf r}) = e^{i{\bf k \cdot r\,^\prime}} + \frac{e^{-3i\pi/4}}{\sqrt{8\pi k}} \,\, \frac{e^{ikr}}{\sqrt{r}} {\cal J},
\en

\noi with

$${\cal J}=\int d^2 r^\prime e^{-ik\frac{{\bf r\cdot r\,^\prime}}{r}} u (r^\prime) \psi^+ ({\bf r\,^\prime}). $$

\bi

Our aim is to obtain the phase shift. To achieve our goal we express the exact
wave function $\psi ({\bf r})$ in terms of a complete set of functions 
$\varphi_\ell ({\bf r})$:

$$\psi^+ (\vec r)= \sum^\infty_{\ell=-\infty} N_\ell \varphi_\ell (r) e^{i\ell\theta} ,$$

\noi the $\varphi (r)$ functions are assumed to have the asymptotic behavior 
(in fact this behavior can be taken as definition of the phase shift $\delta_\ell$)

$$\varphi_\ell (r)\bigg|_{r\to\infty}= \sqrt{\frac{2}{\pi kr}} \cos (kr-\frac{\ell\pi}{2}-\frac{\pi}{4}
+ \delta_\ell) .$$

We also require the two dimensional version of the plane wave expansion in 
terms of ``spherical" waves (states of definite angular momentum) [14]:

$$e^{i k r \cos \theta} = \sum^\infty_{\ell = - \infty} i^\ell J_\ell (kr) 
e^{i\ell\theta}.$$

The angular part of ${\cal J}$ can be explicitly calculated 

$${\cal J}=2 \pi N_\ell (-i)^\ell e^{i \ell \theta}\int r\,^\prime dr^\prime J_\ell(kr^\prime) U (r^\prime) \varphi_\ell (r^\prime) .$$

\noi Furthermore, the asymptotic behavior in both sides of equation (14) leads
to the relation $(x=kr - \frac{\ell\pi}{2} -\frac{\pi}{4})$:

$$N_\ell \cos (x+\delta_\ell) = i^\ell \cos x - i (\frac{\pi}{2}) N_\ell 
\int^\infty_0 r^\prime dr^\prime J_\ell (kr^\prime) U (r^\prime) \varphi_\ell 
(r^\prime) e^{ix} .$$

\noi Solving for each value of the angular momentum $l$ we obtain the  
coefficients $N_\ell$ and an integral expression for the phase shift (compare 
with of Roman [15]):

$$N_\ell = i^\ell e^{i\delta_\ell}$$ 
\be
sin ~\delta_\ell = -\frac{\pi}{2} \int\limits^\infty_0 r^\prime dr^\prime 
J_\ell (kr^\prime) U(r^\prime) \varphi_\ell (r^\prime) .
\en

\noi Below we consider the $1/r^2$ and $\delta({\bf r})$ potentials separately. 

\bi
\bi

\noi {\Large\bf $1/r^2$ potential}

\bi
\bi

For the $1/r^2$ potential the exact solution is $\varphi_\ell(r)=J_\ell (kr)$. 
Substituting this in (15), the phase shift takes the form:

$$sin \delta_\ell =- m\pi\lambda \int^\infty_0 rdr J_\ell (kr) \frac{1}{r^2}
J_{\nu(\ell)} (kr). $$

\noi This integral is tabulated in [14]

$$\int^\infty_0 J_\alpha (x) J_\beta (x) x^{-\gamma} dx= \frac{\Gamma(\lambda)\Gamma(\frac{1}{2} (\nu+\mu-\lambda +1))}{2^\lambda \Gamma(\frac{1}{2} (-\nu+\mu+\lambda+1))\Gamma (\frac{1}{2} (\nu+\mu+\lambda+1)) \Gamma (\frac{1}{2} (\nu-\mu+\lambda +1))}$$

\noi using the reflection formula $\Gamma (z)\Gamma (1-z)=\pi csc (\pi z)$ we 
finally we obtain:

$$\delta_\ell =\frac{\pi}{2}(\ell-\nu(\ell)) +n\pi, \qquad\qquad n=0,1,\ldots$$

\noi the phase shift coincides with relation (8) obtained through the 
asymptotic behavior of the exact eigenfunctions. For completeness we quote the
scattering wave function:

\be
\psi^+ ({\bf r})=\sum^\infty_{\ell=-\infty} e^{i\delta_\ell} i^\ell J_{\nu(\ell)} (kr) e^{i\ell\theta} .
\en

Comparing (16) and (13) we obtain the scattering amplitude:

$$f(\theta)= \frac{1}{\sqrt{2\pi k}}\,\, \sum^\infty_{\ell=-\infty} (e^{2i\delta\ell}-1) e^{i\ell\theta}) .$$

\noi It can be observed that, as scale invariance requires, the differential 
cross section is also energy independent, that is the angular probability of 
scattering and the full cross section are not affected by the energy of the 
incoming particle. 

\bi

\noi {\Large\bf $\delta^2 ({\bf r})$ potential.}

\bi

Our task is to analyze the behavior of the integral representation of the 
phase shift. To this end we notice that, independently of the distribution we 
use to regularize the delta potential, the integral is over the finite
interval (0,{\it a}) and the exact solution $\psi$ is given by (9), therefore:

\be
sin \delta_\ell =-\frac{\pi}{2} \lim\limits_{a\to 0} \int^\infty_0 r^\prime 
dr^\prime J_0 (kr^\prime) \frac{\upsilon (a)}{\pi a^2}d_0 I_0(\kappa r^\prime).
\en

\noi For non vanishing values of the angular momentum, and in the $a\to 0$ 
limit, the integral appearing in (17) is finite whereas in the same limit $v(a)
\cdot d_0 = 0$. Thus, as in the previous section, we conclude the absence of 
scattering for contact potential and $l \not= 0$.

\bi

For the S wave $(l= 0)$ we first consider $v (a)$ given by Eq. (3a). Using
(10b) the phase shift reduces to:

$$sin ~\delta_0 = \frac{\pi}{2} \,\,\frac{b_0}{\ln(\frac{ka_0}{2})} ,$$

\noi which, recalling the comment beneath Eq. (9), can be re-written as:

$$tan ~\delta_0 = \frac{\pi}{2} \,\, \frac{1}{\ln(\frac{k a_0}{2})}.$$

\noi the energy dependence of this phase shift indicates the breaking of scale 
invariance, in agreement with (11), with previous analysis in the literature 
[10] and reproduced in the previous section. It should be clear by now that a 
different result is obtained if instead of (3a) we consider $v(a) =v =$ 
constant. Indeed, in this case, in the $a \to 0$ limit we obtain a vanishing 
phase shift.

\bi
\bi

\noi {\Large\bf SELF ADJOINT EXTENSION}

\bi
\bi

In his contribution to Beg's Memorial Volume, Jackiw proved that the self 
adjoint extension is an alternative approach to the description of the two 
dimensional renormalized delta potential [9]. As emphasized by Jackiw, this 
approach has the advantage -- besides providing a more satisfactory 
mathematical frame -- that it avoids the need to deal with infinite 
quantities, and allow a clear understanding of why the symmetry is broken 
quantum mechanically. On the other hand, in the previous section we argued 
that the conventional approach to the regularized delta potential breaks scale
invariance whereas that a properly regularized delta potential preserves the 
symmetry both at the classical and quantum level and leads to a trivial S 
matrix. It is our purpose in this section to show that the triviality of the S
matrix is also consistent with the self adjoint extension approach. We are not
claiming that the self-adjoint extension treatment is not valid or incorrect, 
we only remark that, under different assumptions, the results of [9] admit a 
different interpretation.

\bi

Consider the radial equation (4). Extracting a $\sqrt{r}$ factor from the wave 
function ({\it i.e.} instead of $\varphi$ it is convenient to introduce the 
function $\tilde\varphi(r) = \sqrt{r} \varphi(r))$ this problem is reduced to 
a one dimensional quantum mechanical system restricted to the half line:

\be
\frac{d^2\tilde\varphi}{dr^2}-\frac{(\ell^2-\frac{1}{4})}{r^2} \tilde\varphi
-2U(r) \tilde\varphi + k^2 \tilde\varphi = 0. 
\en

\noi We begin by considering the self adjoint condition for a free particle 
(notice that this condition is not modified by adding an hermitean potential),

\[ \int^\infty_0 \varphi^*_1 \frac{d^2\tilde\varphi_2}{dr^2} dr= \int^\infty_0
\Bigl(\frac{d^2\tilde\varphi_1}{dr^2}\Bigr)^* \tilde\varphi_2 dr , \]

\noi integrating by parts and assuming that $\varphi (r)$ vanishes at infinity
(to assure normalizability), it follows that the hamiltonian is self-adjoint 
on the set of wave function that satisfies the boundary condition:

\be
\lim\limits_{r\to 0} \Bigl( \varphi^*_1 \frac{d\tilde\varphi_2}{dr} - \Bigl(\frac{d\tilde\varphi_1}{dr}\Bigr)^* \varphi_2 \Bigr)=0 . 
\en

Thus the self-adjoint property is not limited to the operator (the hamiltonian
in this case) but includes also the space of wave functions. When the $r\to 0$ 
limit exist, both for the function and its derivative, (19) is conveniently 
summarized in the condition [16]:

\be
\tilde\varphi^\prime (0) =- c \tilde\varphi (0) , 
\en

\noi in the nomenclature of mathematical physics [9,17] we say that the free 
hamiltonian admits a one parameter family of self adjoint extensions labeled 
by the real parameter $c$. The physical interpretation of these boundary 
conditions is as follows. The function $e^{-ikr} (e^{ikr})$ is a plane wave 
moving to the left (right) with momentum $k >0$, {\it i.e.} is an outgoing 
(incoming) wave of momentum {\bf k}. Clearly these are not square integrable 
functions; however we ignore that since we are only interested in its behavior
near the origin. Neither $e^{-ikr}$ nor $e^{ikr}$ belong to the space of 
functions leading to a self-adjoint free hamiltonian. At this point it is 
convenient to introduce the wave function:

\be
\chi (r) = e^{-ikr} + \alpha e^{ikr}
\en

\noi If $\alpha = \frac{ik-c}{ik+c}$, then $\chi (r)$ satisfies the boundary 
condition (20). Thus the free hamiltonian together with the boundary condition
(20) generate the dynamics in which a plane wave of momentum $k$ is scattered. 
Different self-adjoint extensions (different $c`s$) produce different $(\alpha`s)$, {\it i.e.} different self adjoint extensions correspond to different 
physics (potentials).

\bi

The self-adjoint extension method can be used in different ways, in particular
it can be used to test the symmetry after quantization. Indeed if $\psi(r)$ is
a wave function that satisfies the boundary condition (20), if ${\bf D}$ is 
the generator of a symmetry transformation, and if the symmetry is not broken 
by the quantization process, then $\psi`(r) ={\bf D}\psi(r)$ must also satisfy
the boundary condition (20). Below we apply this criteria to the $\delta^2 ({\bf r})$ potential.

\bi

Following Jackiw [9],  to apply the self adjoint method we consider the exact,
 external S wave ($l = 0$) wave function which for the $\delta^2 ({\bf r})$ 
potential have the following behavior at the origin (recall that $\tilde\varphi (r) = \sqrt{r} \varphi(r)$:

\bi

\be
\tilde\varphi_0(r)=b_0 \sqrt{r} \Bigl( J_0 (kr)+\frac{c_0}{b_0} Y_0 (kr)\Bigr)
\en
\[\to \left[ \sqrt{r} \Bigl(1+ \frac{2\gamma}{\pi} tan \delta_0 \Bigr) + \frac{2 tan \delta_0}{\pi} \sqrt{r} \log \Bigl(\frac{kr}{2} \Bigr) \right] \overrightarrow{_{_{r\to 0}}} 0. \]

\bi

Eventhough the wave function vanishes at the origin, since the derivative $\tilde\varphi^\prime (r)$ is singular at that point:

\[ \tilde\varphi^\prime_0 (r) \overrightarrow{_{_{r\to 0}}} \frac{1}{2\sqrt{r}}
\Bigl(1+ \frac{4 tan \delta_0}{\pi} \Bigl( 1+\frac{\gamma}{2} \Bigr)\Bigr) + \frac{1}{\sqrt{r}} \frac{tan \delta_0}{\pi} \log \Bigl( \frac{kr}{2} \Bigr) , \]

\bi

\noi then, for consistency, we consider condition (19) instead of (20). For
$\tilde \varphi_1$ and $\tilde \varphi_2$ we take free particle solutions $\varphi^i_0 = b^i_0\, (J_0 (kr)+ tan \delta^i_0 Y_0 (kr))$ where $\delta^i_0$ 
stands for the phase shift of the $i-esim$ solution (see comment beneath eq. 
(9)). It is straightforward to show that the selfadjointedness condition (19)
requires $\delta^1_0 =\delta^2_0$, {\it i.e.} that the hamiltonian is 
selfadjoint on the class of functions with the same phase shift. This 
characteristics can be used to test the symmetry after quantization. If the
quantization process preserves the symmetry, then the symmetry generator must
not change the ratio of the $J_0, Y_0$ contributions in (22). Given {\bf D} $Ht
+\frac{i}{2} (r\partial r+1)$, we see that only the $r\partial r$ term can 
change the ratio under consideration. Applying this criterium to (22), we
conclude that scale invariance survives the quantization process only for 
$\delta =0$. Thus, the self adjoint extension approach is consistent with the 
result obtained by explicit evaluation of the phase shift (see also [9]). 
 
\bi
\bi

\noi {\Large\bf SUMMARY}

\bi

In this paper we considered the quantum mechanics of scale invariant potentials
$(1/r^2$ and $\delta^2 ({\bf r})$ potentials). We have shown that a scale 
invariant regularization of the $\delta^2({\bf r})$ potential leads to a 
trivial S matrix. The triviality of the S matrix is not related to the 
strength of the potential, in fact the scale invariant regularization has a 
stronger singularity at the origin that the regularization leading to a non
trivial S matrix. We conclude that scale invariance survives the process of
quantization and that the symmetry breaking of scale invariance reported in 
the literature for the $\delta^2 ({\bf r})$ potential is an example of 
explicit breaking and not an anomaly. We have indicated how the same result 
can be consistently obtained within the self-adjoint extension approach.

\bi

One is tempted to extrapolate our conclusion to quantum field theory. At that
level the question is wheter the renormalization is part of the quantization
or not; if it is not, then dimensional transmutation could be considered an
explicit breaking and not an anomaly!!! 

\newpage
 \begin{center} {\large \bf REFERENCES}
\end{center}
\small
\begin{itemize}
\item[1.-] L.D. Landau, E.M., Lifshitz ``Mechanics". Ed. Riverte (1978). 
\item[2.-] J.M. Levy Leblond, Am. J. Phys. (1971) 502. 
\item[3.-] See for example J. Govaerts ``Hamiltonian Quantisation and  
	   Constrained Dynamics". Leuven University Press (1991). 
\item[4.-] B. Holstein Am. J. Phys. 61 (1993) 142. 
\item[5.-] G.V. Dunne, R. Jackiw and C.A. Trugenberger, Phys. Rev. D41 (1990)
	   661. 
\item[6.-] A. Cabo and J.L. Lucio M. Phys. Lett. A 219, (1996) 155.
\item[7.-] B. Holstein, Am. J. Phys. 61 (1993) 142, C. Manuel R. and Tarrach. 
	   Phys. Letts. B 328 (1994) 113. See also ref. 9. 
\item[8.-] M.A.B. Beg and R. Furlong, Phys. Rev. D31, 1370 (1985). 
\item[9.-] R. Jackiw, in M.A.B. Beg Memorial Volume, eds. A. Ali and P.
	   Hoodbhoy (World Scientific, Singapore, 1991).
\item[10.-] P. Godzinsky and R. Tarrach. Am. J. Phys. 59 (1991) 70, see also
	   L.R. Mead and J. Godines, Am. J. Phys. 59, (1991) 935, C. Manuel and
	   R. Tarrach, Phys. Letts. B 328, (1994) 113.
\item[11.-] S.K. Adhikari and T. Frederico. Phys. Rev. Letts. 74, (1995) 4572
\item[12.-] See first paper of ref. 10 and references therein. 
\item[13.-] A. Cabo, J.L. Lucio M. and M. Napsuciale. Ann. Phys. 244 (1995) 1
	    and references therein.
\item[14.-] M. Abramowitz and I. Stegun, Handbook of Mathematical Functions 
	    Dover, New York (1970).
\item[15.-] P. Roman. ``Advanced Quantum Theory". Adison Wesley, 
	    Massachussetts (1965). 
\item[16.-] E. D'Hoker and L. Vinet Commun. Math. Phys. 97 (1985) 391.
\item[17.-] Reed M., Simon B.: Fourier Analysis and Selfadjointedness New York,
	   Academic Press (1975).	    
\end{itemize} 
\end{document}